\documentstyle[12pt,titlepage,epsfig,floatfig]{article}
\begin{document}
\pagestyle{empty}
\baselineskip=0.212in
\renewcommand{\theequation}{\arabic{section}.\arabic{equation}}
\renewcommand{\thefigure}{\arabic{section}.\arabic{figure}}
\renewcommand{\thetable}{\arabic{section}.\arabic{table}}

\initfloatingfigs

\begin{flushleft}
\large
{SAGA-HE-98-96  
   \hfill May 8, 1996}  \\
\end{flushleft}
 
\vspace{2.5cm}
 
\begin{center}
 
\LARGE{{\bf Parton Distributions in Nuclei:}} \\
\vspace{0.3cm}

\LARGE{{\bf Overview and Prospect}} \\
 
\vspace{1.5cm}
 
\Large
{S. Kumano $^*$ }         \\
 
\vspace{0.8cm}
  
\Large
{Department of Physics}         \\
 
\vspace{0.1cm}
 
\Large
{Saga University}      \\
 
\vspace{0.1cm}

\Large
{Saga 840, Japan} \\

\vspace{2.0cm}
 
\large
{Invited talk given at the Workshop on} \\

\vspace{0.3cm}

{``New Developments in QCD and Hadron Physics"} \\

\vspace{0.7cm}

{Kyoto, Japan, Jan.22-24, 1996 (talk on Jan.23, 1996)}  \\
 
\end{center}
 
\vspace{1.5cm}

\vfill
 
\noindent
{\rule{6.cm}{0.1mm}} \\
 
\vspace{-0.2cm}
\normalsize
\noindent
{* Email: kumanos@cc.saga-u.ac.jp. 
   Information on his research is available}  \\

\vspace{-0.6cm}
\noindent
{at http://www.cc.saga-u.ac.jp/saga-u/riko/physics/quantum1/structure.html} \\

\vspace{-0.6cm}
\noindent
\normalsize
{or at ftp://ftp.cc.saga-u.ac.jp/pub/paper/riko/quantum1.} \\

\vspace{1.0cm}

\vspace{-0.5cm}
\hfill
{to be published in Soryushiron Kenkyu (Kyoto)}

\vfill\eject
\pagestyle{plain}
\begin{center}
 
\Large
{Parton Distributions in Nuclei: Overview and Prospect} \\
 
\vspace{0.5cm}
 
{S. Kumano $^*$}             \\
 
{Department of Physics, Saga University}    \\
 
{Honjo-1, Saga 840, Japan} \\

\vspace{0.7cm}

\normalsize
Abstract
\end{center}
\vspace{-0.30cm}

We give a brief overview of nuclear parton distributions.
First, the EMC effect is discussed together with 
possible interpretations such as nuclear binding and
$Q^2$ rescaling. Next, we explain shadowing descriptions:
vector-meson-dominance-type and parton-recombination models.
Nuclear dependence of $Q^2$ evolution should be interesting 
in testing whether or not DGLAP equations could be applied
to nuclear structure functions.
Status of nuclear sea-quark and gluon distributions is discussed.
The structure function $b_1$, which will be measured at HERMES
and possibly at ELFE, could shed light on a new aspect of 
high-energy spin physics.

\setcounter{page}{1}
\section{{\bf Introduction}}\label{INTRO}
\setcounter{equation}{0}
\setcounter{figure}{0}

From the early SLAC experiments on electron scattering, nuclear targets
have been used just as a tool for measuring nucleon
structure functions. Nuclear corrections are not taken seriously
in 1970's except for apparent nucleon-Fermi-motion corrections.
Nucleons are bound in a nucleus with average binding energy
8 MeV and typical momentum 200 MeV.
These are fairly small compared with reaction
energy of several GeV or more in electron or muon
scattering. For this reason, nuclear environment
was not expected to alter nucleon properties significantly.
Even though signature of nuclear modification in the $F_2$ structure
function is found in the SLAC data in 1970's, they are not 
studied extensively. On the other hand, the European Muon Collaboration 
(EMC) took the topic rather seriously and published the first paper 
in 1983 \cite{EMC83}.  The nuclear modification of $F_2$ is now
known as the (old) EMC effect. 

In the mid 1980's, there are many theoretical publications
on this topic, in particular on medium and large $x$ physics. 
We cannot list all of the ideas in this paper.
A conservative one is to interpret it in term of nuclear binding
and Fermi motion \cite{AKV85}. 
On the other hand, there is another extreme model, $Q^2$ rescaling 
\cite{RESCALE}. 
In contrast to the binding model in which bound nucleon $F_2$
is the same as the free one, it is modified 
due to nuclear environment. The details of these models are discussed
in section \ref{EMCF2}.

From the late 1980's, nuclear $F_2$ structure functions 
in the shadowing region ($x<0.1$)
are investigated in detail experimentally and theoretically.
In particular, accurate New Muon Collaboration (NMC) data
make it possible to test theoretical shadowing models.
The traditional idea of describing the shadowing is
a vector-meson-dominance (VMD) model. The virtual
photon transforms into vector-meson states, which
interact with the target nucleus.
The VMD contribution decreases as $1/Q^2$ at large $Q^2$,
so various extensions of the model 
($q\bar q$ continuum, Pomeron exchange) are studied.
On the other hand, there exists an infinite-momentum interpretation
in terms of parton recombinations, which are parton interactions
from different nucleons in the nucleus.
Shadowing phenomena are discussed in section \ref{SHADOW}.
Comments on sea-quark and gluon distributions are also given.

Bjorken scaling works roughly for structure functions, and
$Q^2$ dependence is a small effect. 
It is inevitably difficult to find nuclear effects within
the small $Q^2$ variation. 
However, the nuclear dependence is recently found in a NMC experiment 
\cite{NMCSNC} by measuring tin and carbon $Q^2$ variation
differences.
The data are interesting in testing whether or not
DGLAP equations could be applied to the nuclear case.
This topic is discussed in section \ref{Q2}.

High-energy spin physics becomes a very popular topic
since the EMC discovery of proton-spin ``crisis".
In order to investigate a new field of spin physics,
structure of the spin-one deuteron is currently under investigation 
by HERMES.
It is expected that a new spin field is explored in the near future.
Major features of the structure function $b_1$
are discussed in section \ref{B1}.

\section{{\bf Nuclear modification of $\bf F_2$ at medium $\bf x$}}
\label{EMCF2}
\setcounter{equation}{0}
\setcounter{figure}{0}

\begin{floatingfigure}{7.0cm}
   \begin{center}
      \mbox{\epsfig{file=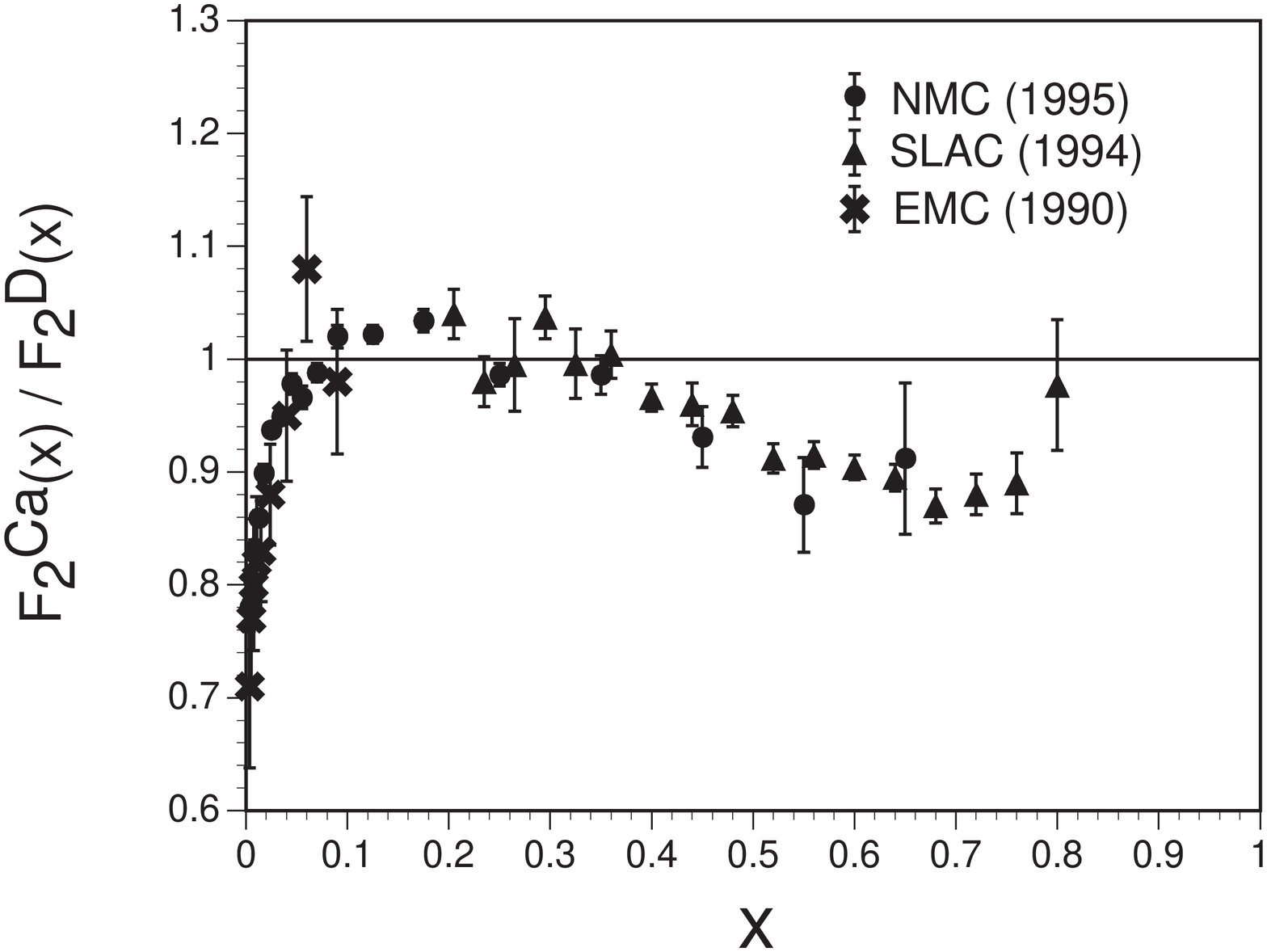,width=5.0cm}}
   \end{center}
 \vspace{-0.8cm}
\caption{\footnotesize Nuclear modification $F_2^{Ca}/F_2^D$.} 
\label{fig:F2CA-D}
\end{floatingfigure}
\quad
High-energy electron or muon scattering has been used for probing internal
structure of nucleons or nuclei.
Its cross section is written in terms of two structure functions
$F_1$ and $F_2$. $F_1$ is related to the transverse cross section
for the virtual photon, and $F_2$ is to the transverse and longitudinal
ones. If higher-order $\alpha_s$ corrections are neglected,
they are related by the Callan-Gross relation $2xF_1=F_2$. 
Therefore, we discuss only the $F_2$ structure function.
In parton model, the cross section is described by the incoherent
summation of scattering cross sections from individual quarks:
\begin{equation}
F_2 (x) = \displaystyle{\sum_i} e_i^2 \ x \ [q_i (x) +\bar q_i(x)]
\ \ \ .
\end{equation}

We discuss how the structure function $F_2$ is modified in nuclei.
The details of the modification is studied experimentally 
for several nuclei. For example, SLAC, EMC, and NMC data \cite{CAEXP} 
for the calcium-deuteron $F_2$ ratio are shown in Fig. \ref{fig:F2CA-D}.
In the medium $x$ region ($x\sim$0.5), 
the ratios are approximately 10\% smaller
than unity, and it tends to increase at large $x(>0.8)$.
In the region $x\approx 0.2$, the modification is positive,
and the phenomenon is called ``antishadowing" in contrast with
shadowing at small $x$. The shadowing part is discussed
in section \ref{SHADOW}.

In describing nuclear structure functions, convolution
formalism is usually used. 
A parton distribution in a nucleus
is given by a hadron (nucleon, pion, and etc.) momentum distribution
convoluted with a parton distribution in the hadron.
In terms of light-cone momenta, it is written as
\begin{equation}
f_{a/A}(x_A,Q^2) = \sum_T \int_{x_A}^1 dy_A 
                    f_{a/T}(x_A/y_A,Q^2) f_{T/A}(y_A)
\ \ \ ,
\end{equation}
where $x_A$ is defined by $x_A=Q^2/2M_A\nu$ and 
$y_A$ is the light-cone momentum fraction $y_A=p_T^+/p_A^+$.
In order to explain the modification of the quark distribution
in a nucleus [$f_{a/A}(x_A,Q^2)$], 
we may ascribe it to modification of the quark distribution
in a hadron [$f_{a/T}(x_A/y_A)$] or to modification
of the hadron distribution in the nucleus [$f_{T/A}(y_A)$].

A standard approach in describing nuclei in the low-energy region
is to start from information on nucleon-nucleon interactions, which are
obtained by NN scattering experiments and deuteron properties.
Then, the knowledge is applied to nuclear structure by
taking into account many-body corrections.
In this sense, it is a ``proper" approach to start from
an assumption that the structure function $F_2$ of a bound nucleon
is the same as the free nucleon $F_2$.
This way of interpreting the EMC effect is a nuclear binding model
\cite{AKV85}. 
A nuclear tensor, which appears in calculating a lepton-nucleus 
cross section, is given by the convolution:
$W_{\mu\nu}^A(P,q)={\displaystyle \int} d^4p S(p) W_{\mu\nu}^N(p,q)$.
The spectral function $S(p)$ is the nucleon momentum
distribution in the nucleus.
If we consider the simplest case of a shell model,
it is given by single particle wave functions
and binding energies of nucleons:
$S(p)={\displaystyle \sum_i} |\phi_i(\vec p)|^2
       \delta (p_0-m_N-\epsilon_i)$.
Recoil energy is neglected in the expression.
From these equations, the nuclear $F_2$ is
\begin{equation}
F_2^A(x,Q^2)=\sum_i \int dz f_i(z) F_2^N(x/z,Q^2)
\ \ \ ,
\end{equation}
where $x$ is the Bjorken variable $x=Q^2/2m_N\nu$, and
$f_i(z)$ is the light-cone momentum distribution of the nucleon $i$:
\begin{equation}
f_i(z)= \int d^3 p \ z \ \delta 
        \left ( z -{{p\cdot q}\over {m_N\nu}} \right )
        | \phi_i(\vec p)|^2
\ \ \ .
\end{equation}
The momentum fraction $z$ is given by
$z=p\cdot q/m_N\nu=1-|\epsilon_i|/m_N+\vec p\cdot \vec q/m_N\nu
  \approx 1.00 - 0.02\pm 0.2$ for a typical nucleon.
So, the function $f(z)$ is peaked at 0.98. 
If we replace $f(z)$ by a delta function $\delta (z-0.98)$
for simplicity, the nuclear $F_2$ becomes
$F_2^A(x,Q^2)\approx F_2^N(x/0.98,Q^2)$. 
For example at $x$=0.60, we have $F_2^N(x=0.61)/F_2^N(x=0.60)=0.88$ 
and the 10\% EMC effect can be explained at medium $x$.

On the other hand, there exist other extreme interpretations
in terms of modification of nucleon itself.
A well known idea is the $Q^2$ rescaling model.
It seems that the nucleon modification makes sense
if we consider that the average nucleon separation in a nucleus
is about 2 fm, which is almost equal to the nucleon diameter.
Therefore, it is no wonder that multiquark systems
other than the nucleon are formed.
If such multiquark hadrons are created, a confinement size
for quarks should be changed. 
Then the quark momentum distributions are modified according
to the size change.
Using this kind of simple picture at a hadronic
scale $\mu^2$, we calculate distributions 
at large $Q^2$, where experimental data are taken,
by evolving distributions from $\mu^2$. 
We find that a nuclear $F_2$ is related 
to the nucleon one simply by scaling $Q^2$:
\begin{equation}
F_2^A(x,Q^2)=F_2^N(x, \xi_A Q^2)
\ \ \ ,
\end{equation}
with the rescaling parameter
$\xi_A=(\lambda_A^2/\lambda_N^2)^{\alpha_s(\mu_A^2)/\alpha_s(Q^2)}$,
where $\lambda$ is the confinement radius for a quark.

The binding and rescaling models seem very different interpretations.
How shall we understand the situation? A possible explanation is
in terms of factorization scale independence \cite{RESCALE}. 
For example in the operator-product-expansion case,
the renormalization scale $\mu$ separates long-distance physics
and short-distance one. However, final results should not depend
on the arbitrary human factor $\mu$.
Because the bound structure function and the nucleon momentum 
distribution are {\it not} separately observables, the same discussion
could be valid in our case. The nuclear structure function,
which is a physics observable, should not depend on a factorization 
scale $\mu$ which separates it into the quark-distribution
part and the hadron-distribution one. Choosing a nuclear-dependent 
scale $\mu=\mu_A$, we obtain the rescaling model. On the other hand,
the binding model corresponds to the nuclear-independent scale.
Using this factorization-scale independence, we may relate 
the two different descriptions.
Moments of a nonsinglet distribution are written as
$M_n^{q/A}=M_n^{q/N} M_n^{N/A}$ by
Mellin-transformation properties.
As a simple example of the binding model, a Fermi-gas model
is employed. 
Relations between the two descriptions are found as
$\bar\epsilon/m_N=-\gamma_2^{NS}\kappa_A/2\beta_0$
and $k_F^2/m_N^2=5(2\gamma_2^{NS}-\gamma_3^{NS})\kappa_A/2\beta_0
          +O(\bar\epsilon^2/m_N^2)$, where $\kappa_A$ is defined by
$1-\kappa_A=\alpha_s(\xi_A Q^2)/\alpha_s(Q^2)$.
The quantities $\bar\epsilon$ and $k_F$ are 
the average binding energy and the Fermi momentum, and
$\gamma_n^{NS}$ is the nonsinglet anomalous dimension.
There exists a correspondence between the binding model and 
the rescaling, so that we may view the rescaling as an effective
description of nuclear medium effects.
In fact, it is shown in a simple quark model that effects of
nucleon interactions could be effectively described by
a nucleon size change in the nuclear medium \cite{KM},
although it is perhaps too simple to explain various
nuclear phenomena.
Comparison of the rescaling results with the data is
discussed in section \ref{SHADOW}.

\section{{\bf Nuclear shadowing}}\label{SHADOW}
\setcounter{equation}{0}
\setcounter{figure}{0}
\setcounter{table}{0}

Nuclear modification of $F_2$ at small $x$ is negative
as shown in Fig. \ref{fig:F2CA-D}, and it is known as shadowing
phenomena.
A traditional way of describing the shadowing is in terms of
vector-meson-dominance (VMD) model.
The virtual photon transforms into vector meson states,
which interact with the target.
Because a propagation length of the vector mesons is given by
$\lambda=1/|E_V-E_\gamma|\approx$0.2/x fm, it exceeds 2 fm
at $x<$0.1 and multiple scattering occurs.
The vector mesons interact predominantly in the surface region,
and internal nucleons are ``shadowed" by the surfaces ones.
This description is not enough for explaining observed
small $Q^2$ dependence. So various extensions
of this model are studied. For example, including $q\bar q$
continuum, we have the nuclear $F_2$ \cite{VMDTYPE}
\begin{equation}
F_2^A(x,Q^2) = {{Q^2} \over \pi } \int dM^2
               {{M^2 \Pi \left( M^2 \right)} \over 
                {\left( M^2 +Q^2 \right) ^2 }} \sigma_{VA}
\ \ \ ,
\end{equation}
where $\sigma_{VA}$ is the vector-meson nucleus interaction
cross section, and $\Pi \left( M^2 \right)$ is the spectral
function $\Pi \left( M^2 \right)=\sigma(e^+ e^- \rightarrow hadrons)/
                                 \sigma(e^+ e^- \rightarrow \mu^+ \mu^-)
              = \ vector\ mesons \ + \ q\bar q\ continuum$.

Another interpretation of shadowing is in term of Pomeron exchange
\cite{BGNPZ}. In the case of diffractive scattering, 
the target is thought to be remain intact so that only vacuum 
quantum number, ``Pomeron", could be exchanged. 
The Pomeron ($\cal P$) structure function is
defined by the diffractive cross section:
$F_{2,{\cal P}} = Q^2/(4\pi^2\alpha) \sigma_{\gamma^* {\cal P}}$.
Pomeron contribution to the nuclear $F_2$ from double diffractive
scattering is given by a convolution of the Pomeron $F_2$ with
its light-cone momentum distribution:
$\delta F_{2}(x) = {\displaystyle \int} 
                    dy f_{\cal P} (y) F_{2,{\cal P}}(x/y)$.
The variable $y$ is the momentum fraction carried by 
the Pomeron $y=k\cdot q/p\cdot q$.
The VMD contribution is compared with the Pomeron result
in \cite{MT}. Both shadowing results are of the same order of 
magnitude at small $x$ ($\approx$0.01) and at $Q^2$=4 GeV$^2$.

The above shadowing ideas are in the target rest frame.
There is an explanation in an infinite momentum frame
in terms of parton recombinations \cite{PR,SK}.
Both descriptions are supposedly equivalent; however, there is
no explicit proof at this stage.
The recombinations are parton interactions
from different nucleons. They become dominant at small $x$
with the following reason. 
The average longitudinal nucleon separation in a Lorentz contracted 
nucleus is $L=(2\ fm)M_A/P_A=(2\ fm)m_N/p_N$.
The longitudinal localization size of a parton with momentum  $xp_N$
is $\Delta L=1/xp_N$.
In the $x$ region $x<0.1$, $\Delta L$ becomes larger than $L$, so
that partons from different nucleons could interact significantly.
For example, a $p_1 p_2\rightarrow p_3(x)$ recombination effect
on the $p_3$ parton distribution is
\begin{equation}
\Delta p_3 (x) = {{9A^{1/3}\alpha_s} \over{2R_0^2 Q^2}} \int dx_1 dx_2
p_1(x_1) p_2(x_2) \Gamma_{p_1p_2\rightarrow p_3} (x_1,x_2,x) 
\delta (x-x_1-x_2)
\ \ \ ,
\label{eqn:RECOM}
\end{equation}
where the factor $A^{1/3}$ is the number of nucleons in the longitudinal
direction, $p_1 p_2\rightarrow p_3$ cross section is proportional to
$\alpha_s/Q^2$, and $R_0$ is given by $R_0\approx$1 fm.
The parton-fusion function $\Gamma_{p_1 p_2\rightarrow p_3}$
indicates a probability of the fusion process $p_1 p_2\rightarrow p_3$.
Because it is opposite to the splitting, 
$\Gamma_{p_1 p_2\rightarrow p_3}$ is related to the Altarelli-Parisi
splitting function by
$\Gamma_{p_1 p_2 \rightarrow p_3}(x_1,x_2,x_3) =
 (x_1 x_2/x_3^2) P_{p_1 \leftarrow p_3} (x_1/x_3)
 C_{p_1 p_2 \rightarrow p_3}$, where $C_{p_1 p_2 \rightarrow p_3}$
is the color-factor ratio 
$C(p_1 p_2 \rightarrow p_3)/C(p_3\rightarrow p_1 p_2)$.
From Eq. (\ref{eqn:RECOM}), the recombination shadowing has
typical $A^{1/3}$ dependence. Furthermore, the $1/Q^2$ dependence
indicates that recombinations are higher-twist effects.
Because of this $1/Q^2$ factor, the magnitude of the shadowing
depends much on $Q^2$ at which the recombinations are calculated.

\begin{floatingfigure}{7.0cm}
   \begin{center}
      \mbox{\epsfig{file=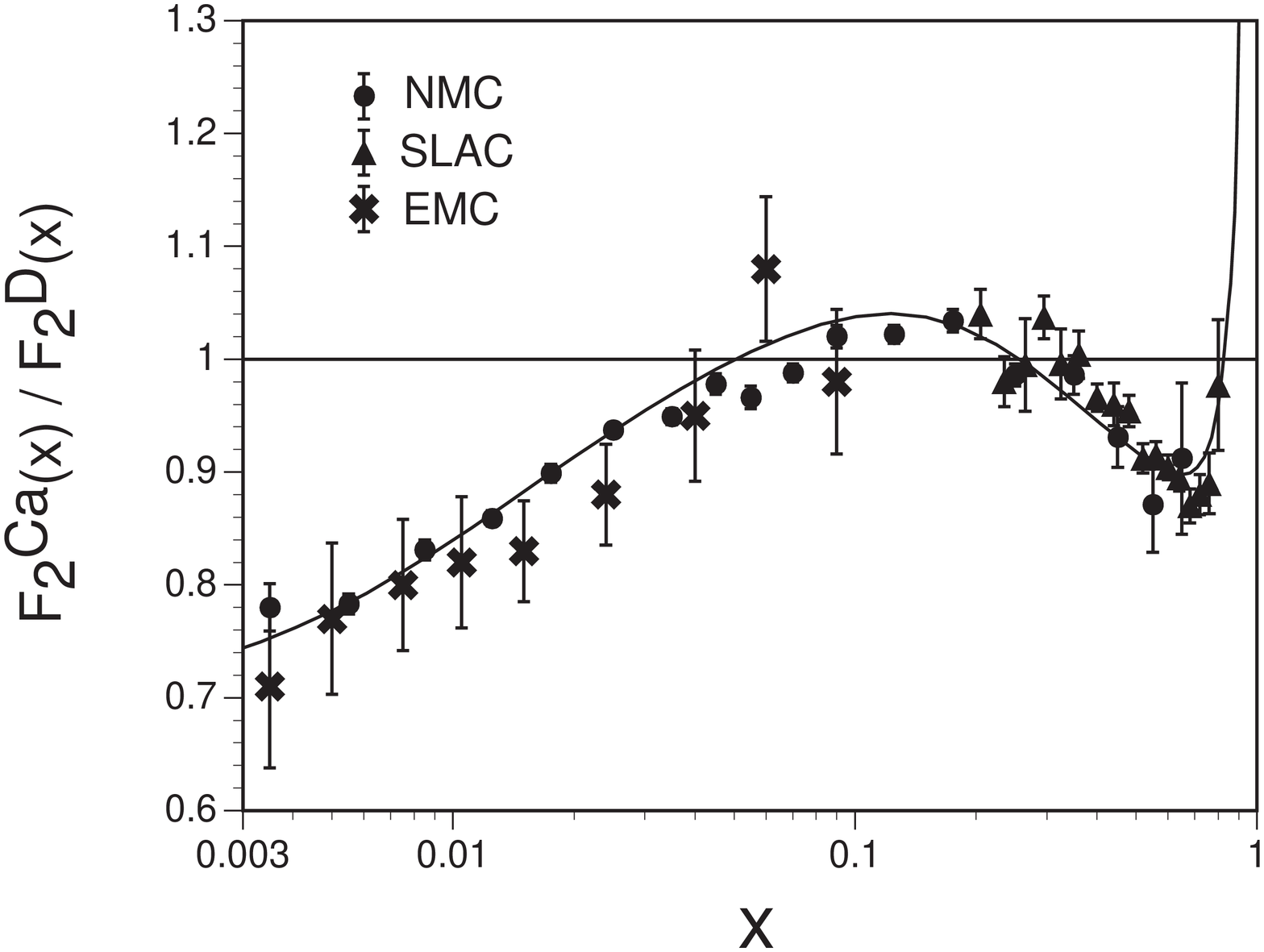,width=5.0cm}}
   \end{center}
 \vspace{-0.8cm}
\caption{\footnotesize Comparison of theoretical results with
                       the data $F_2^{Ca}/F_2^D$.} 
\label{fig:F2CA}
\end{floatingfigure}
\quad
As an example of theoretical results, we show the nuclear modification
$F_2^{Ca}/F_2^D$ in a model with the recombination and the rescaling 
mechanisms \cite{SK}. 
Both contributions are calculated at $Q_0^2$, then
obtained distributions are evolved to the ones at larger $Q^2$.
The initial point $Q_0^2$ is considered as a parameter in the model,
and it is fixed by fitting the NMC shadowing data for the calcium
($Q_0^2$=0.8 GeV$^2$). The parton distributions at $Q^2$=5 GeV$^2$
are calculated by the DGLAP evolution equations, and they are compared
with the data in Fig. \ref{fig:F2CA}.
Although $Q_0^2$ is an adjustable parameter, the model can explain
the nuclear modification from very small $x$ to large $x$.
It is interesting to note that the antishadowing part at $x\approx$0.2 
is explained by competition between the recombination and the rescaling
effects and that the ``Fermi motion" part is described by 
the quark-gluon recombinations.

\begin{table}[h]
\begin{center}
\begin{tabular}{||l|l|l||} \hline
\               ~&~ $x<0.15$          ~&~ $x>0.15$                ~\\ \hline
\ valence quark ~&~  ?                ~&~ $\star\star\star\star$  ~\\
\ sea quark     ~&~ $\star\star\star$ ~&~ $\star\star$            ~\\
\ gluon         ~&~ $\star$ ?         ~&~ $\star$ ?               ~\\ \hline
\end{tabular}
\end{center}
\caption{\footnotesize Status of nuclear parton distributions.}
\label{tab:STATUS}
\end{table}

\begin{figure}[h]
\parbox[c]{0.46\textwidth}{
   \begin{center}
       \epsfig{file=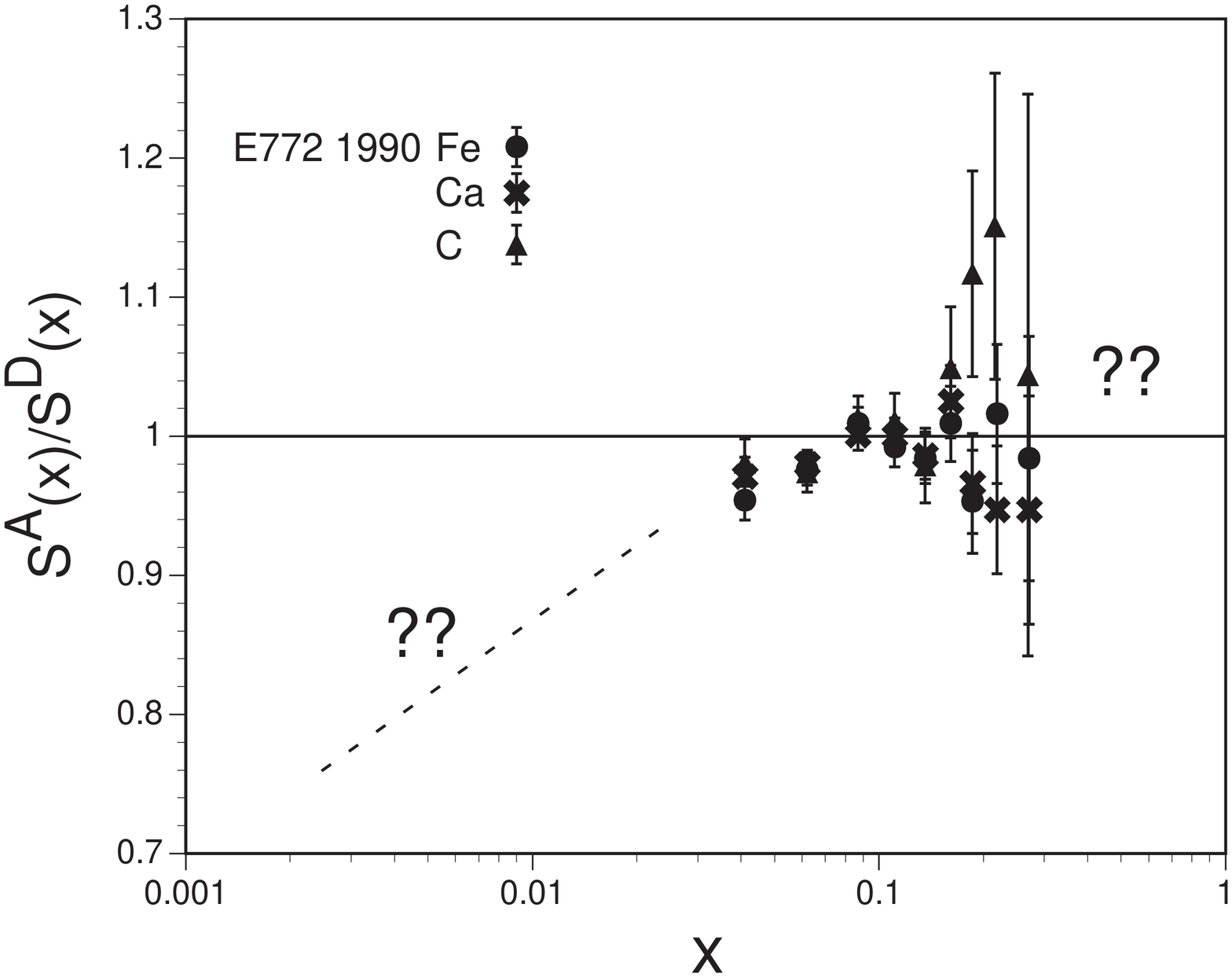,width=5.0cm}
   \end{center}
\vspace{-0.5cm}
       \caption{\footnotesize Sea-quark distribution in Drell-Yan process.}
       \label{fig:SEA}
}\hfill
\hspace{0.5cm}\vspace{0.5cm}
\parbox[c]{0.46\textwidth}{
   \epsfig{file=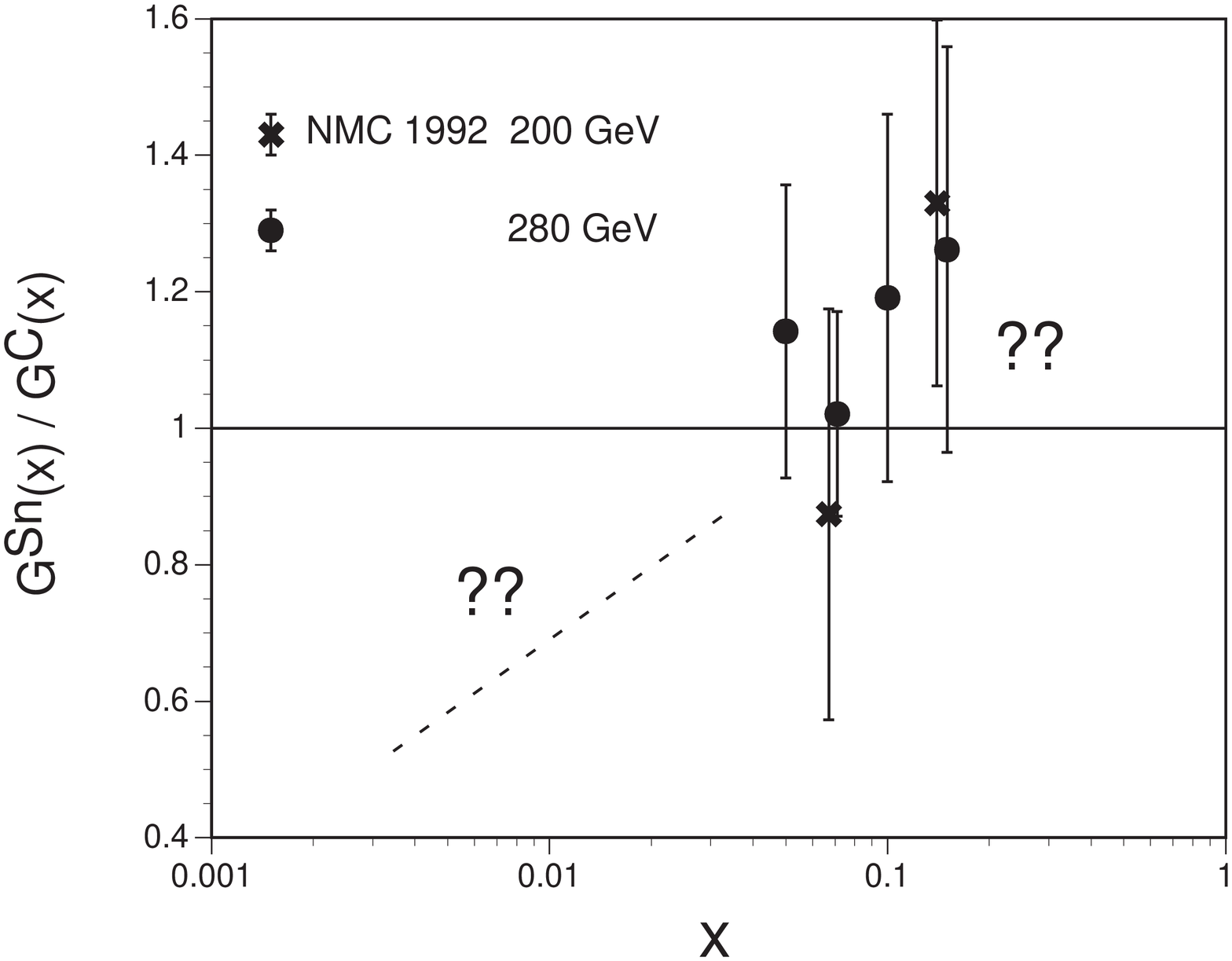,width=5.0cm}
\vspace{0.0cm}
   \caption{\footnotesize Gluon distribution in muon-induced J/$\psi$
                          production.}
   \label{fig:GLUE}
}
\end{figure}

Current status of nuclear parton distributions are summarized in 
Table \ref{tab:STATUS}. The $F_2$ structure function at medium (small) $x$
is essentially the valence (sea) quark distribution. So we have good
information on the valence-quark distribution at medium $x$ and 
the sea-quark one at small $x$ in several nuclei. 
The sea-quark distribution is also investigated in Drell-Yan processes.
The Fermilab Drell-Yan experiment \cite{E772} did not measure the shadowing
region as shown in Fig. \ref{fig:SEA}. 
We hope that future measurements at Fermilab
and at RHIC probe the shadowing region at very small $x$.
The iron data in Fig. \ref{fig:SEA} are often quoted for concluding
that there is little modification in the sea. However, it is not
very obvious by looking at other nuclear data. 
Accurate $A$ dependent data, as well as theoretical studies,
are necessary for finding the details of sea-quark behavior.
Because the sea-quark distribution dominates $F_2$ at small $x$,
behavior of valence quarks at small $x$ is not known. However,
it could be an interesting topic in testing various shadowing models
\cite{KKM}. The valence-quark shadowing could be observed by 
measuring charged pion productions in electron (HERA) or muon scattering.

On the other hand, nuclear gluon distributions are little known.
As an example, we show analysis of
muon-induced J/$\psi$ production data in a color-singlet model. 
Obtained gluon ratios in the tungsten
and carbon are given in Fig. \ref{fig:GLUE} \cite{NMCGLUE}.
At this stage, the errors are too large to indicate nuclear
modification. Furthermore, the gluon shadowing region 
is not measured. We hope to have accurate data
in future, for example at RHIC.

\vspace{-0.8cm}
\noindent

\vspace{-1.0cm}
\section{{\bf Nuclear dependence of $\bf Q^2$ evolution}}\label{Q2}
\setcounter{equation}{0}
\setcounter{figure}{0}

It is known that the Bjorken scaling, which indicates structure functions
are independent of $Q^2$, works approximately. 
Therefore, most studies on the nuclear modification are focused 
on the $x$ dependence. It is difficult to find small nuclear effects
on the scaling violation. However, it became possible to measure
such small effects recently \cite{NMCSNC}. The NMC obtains $Q^2$ evolution
differences in tin and carbon nuclei: 
$\partial(F_2^{Sn}/F_2^C)/\partial(\ln Q^2)$.
It is an interesting topic to investigate whether or not
the DGLAP (Dokshitzer-Gribov-Lipatov-Altarelli-Parisi) equations
could be applied to nuclear structure functions.
In particular, the parton recombinations are predicted to produce
higher-twist effects in the evolution \cite{PR,MK}:
\begin{eqnarray}
& & {\partial \over {\partial t}} \ q_i \left({x,t}\right)\
     =\ \int_{x}^{1}{dy \over y}\ 
      \left[\ \sum_j P_{q_{i} q_{j}}\left({{x \over y}}\right)\ 
           q_j \left({y,t}\right)\ 
       +\  P_{qg}\left({{x \over y}}\right)\ 
        g\left({y,t}\right)\ \right]
\nonumber \\
& & \ \ \ \ \ \ \ \ \ \ \ \ \ \ \ \ \ \ \ \ \ \ \ \ \ \ \ 
    + \ \left( recombination\ terms\ \propto \ 
{{\alpha_s A^{1/3}} \over {Q^2}} \right) \ 
\ \ \ , \\
\label{eqn:AP1}
& & {\partial \over {\partial t}} \ g\left({x,t}\right)\ 
     =\ \int_{x}^{1}{dy \over y}\ 
     \left[\ \sum_j P_{gq_j}\left({{x \over y}}\right)\ 
     q_j \left({y,t}\right)\ 
     +\ P_{gg}\left({{x \over y}}\right)\ 
     g\left({y,t}\right)\ \right]
\nonumber \\
& & \ \ \ \ \ \ \ \ \ \ \ \ \ \ \ \ \ \ \ \ \ \ \ \ \ \ \ 
    + \ \left( recombination\ terms\ \propto \ 
      {{\alpha_s A^{1/3}} \over {Q^2}} \right) \ 
\ \ \ , 
\label{eqn:AP2}
\end{eqnarray}
where the variable $t$ is defined by
$t = -(2/\beta_0) \ln [\alpha_s(Q^2)/\alpha_s(Q_0^2)]$.
The first two terms in Eqs. (4.1) and (4.2) describe 
the process that a parton $p_j$ with the nucleon's momentum fraction $y$ 
splits into a parton $p_i$ with the momentum fraction $x$ and another parton.
The splitting function $P_{p_i p_j}(z)$ determines 
the probability for the parton $p_j$ radiating the parton $p_i$
such that the $p_j$ momentum is reduced by the fraction $z$. 
There are additional recombination contributions in
Eqs. (4.1) and (4.2).

\begin{floatingfigure}{7.0cm}
   \begin{center}
      \mbox{\epsfig{file=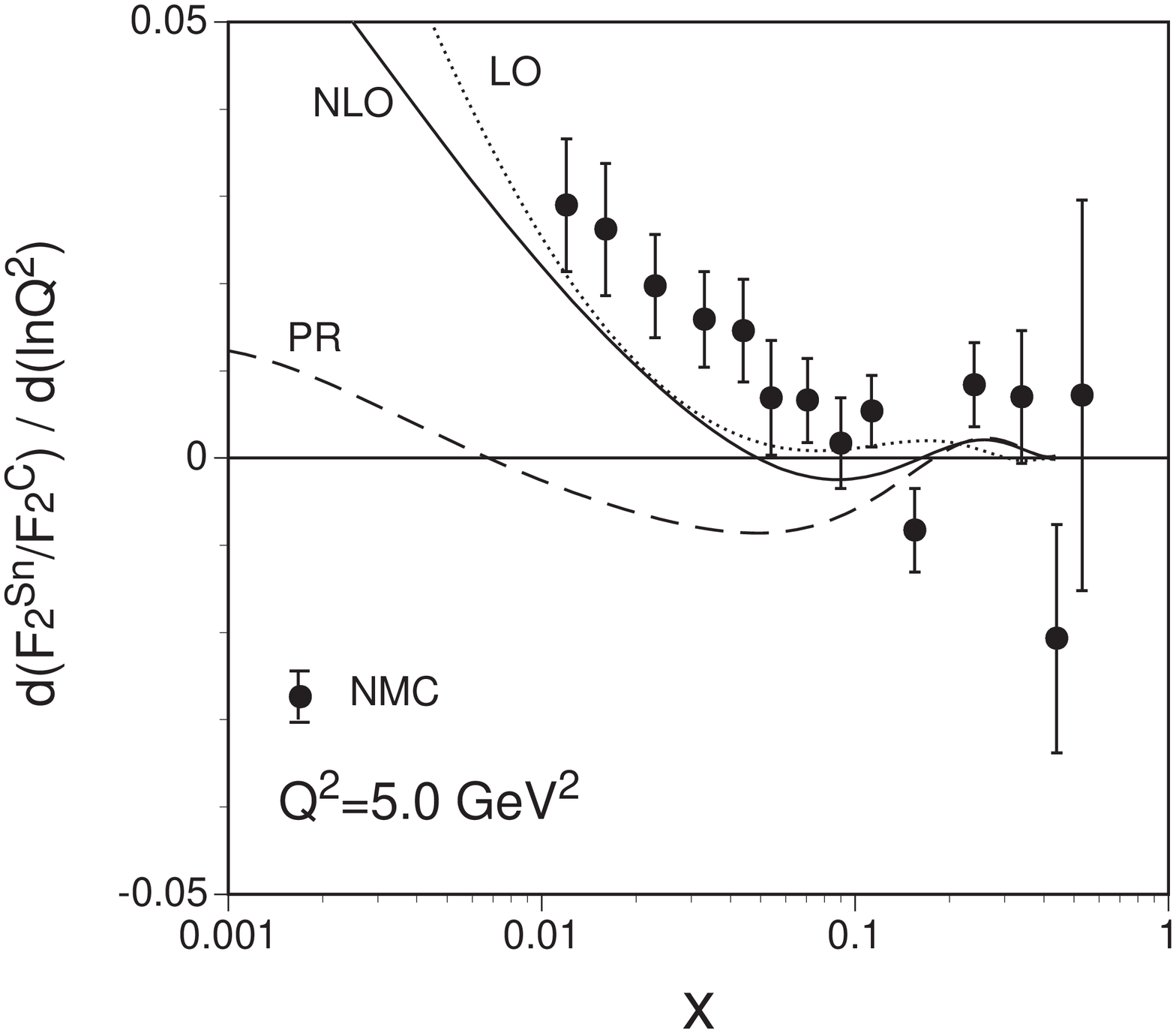,width=5.0cm}}
   \end{center}
 \vspace{-0.8cm}
\caption{\footnotesize Difference of $Q^2$ evolution in tin 
                and carbon nuclei:
                $\partial (F_2^{Sn}/F_2^C)/\partial(\ln Q^2)$.}
\label{fig:SN-C}
\end{floatingfigure}
\quad
There are two possible sources for the nuclear dependence in Eqs. 
(4.1) and (4.2).
First, parton $x$-distributions
are modified in a nucleus so that input distributions
on the right-hand sides are nuclear dependent.
Second, the additional recombination terms have $A^{1/3}$
nuclear dependence. 
We calculate these effects on the nuclear difference
in the $Q^2$ evolution:
$\partial(F_2^{Sn}/F_2^C)/\partial(\ln Q^2)$ \cite{SKMM}.
As initial distributions at $Q_0^2$, 
we use those in the recombination model with the rescaling 
in section \ref{SHADOW}. 
They are evolved to $Q^2$=5 GeV$^2$ by leading-order (LO) DGLAP,
next-to-leading-order (NLO) DGLAP, and parton recombination (PR)
equations with the help of the computer program bf1.fort77 
in Ref. \cite{MK}. 
The results are compared with the NMC data \cite{NMCSNC}
in Fig. \ref{fig:SN-C}.
The DGLAP evolution curves agree roughly
with the experimental data, so that the major nuclear dependence
comes from the nuclear modification in the $x$ distributions.
The PR evolution curve disagrees with the data; however,
it does not mean that the recombination mechanism is ruled out.
There are a few uncertain factors. For example, the input 
higher-dimensional gluon distribution is essentially not known. 
Further refinement of the recombination contributions to the evolution
is necessary. It is nonetheless interesting in the sense that
higher-twist effects could be found in studying the nuclear
$Q^2$ evolution.

\section{{\bf Spin structure of spin-one hadrons}}\label{B1}
\setcounter{equation}{0}
\setcounter{figure}{0}

Spin structure of spin-1/2 hadrons has been investigated
fairly well. In addition,
there is a good possibility to measure a new 
spin structure function for spin-one hadrons in the near future.
As a realistic spin-one target, we have the deuteron.
Studying the deuteron spin structure at high energies,
we gain an insight into a new aspect of spin physics \cite{B1PAPER}.
There are four new structure functions for spin-one hadrons:
$b_1$, $b_2$, $b_3$, and $b_4$.
In the Bjorken scaling limit, the only relevant structure
function is $b_1$ or equivalently $b_2/2x$.
For measuring $b_1$, the electron or muon does not have to be
polarized but we need polarized deuteron. The $b_1$ structure function
is proportional to the tensor combination of the polarized cross sections:
$b_1\propto d\sigma(0)-[ d\sigma(+1)+ d\sigma(-1)]/2$.
In parton model, it is given by
\begin{equation}
b_1 (x) =  {\displaystyle \sum_i} \ e_i^2 \ 
            [\delta q_i(x)+ \delta \bar q_i(x)] 
\ \ \ ,
\label{eqn:B1}
\end{equation}
where $\delta q_i = [ q_i^0 - (q_i ^{+1}+q_i ^{-1})/2 ]/2
           =  {1 \over 2} [{q_i ^0}  (x) 
                            -{q_{i} ^{+1} } (x)]$
with the flavor-i quark distribution $q_i^H(x)$ in $z$ component
of the target spin ($H$). 
Because quark spin does not appear in the above expression,
$b_1$ probes very different spin structure.

We discuss a sum rule for $b_1$ \cite{B1SUM}. 
The integral of Eq. (\ref{eqn:B1}) over $x$ is
related to a tensor combination of
elastic amplitudes by using the parton picture
in an infinite momentum frame.
Next, the elastic helicity amplitudes are written by
charge and quadrupole form factors of the spin-one hadron.
In the case of tensor combination, the charge form factor
cancels out and the remaining term is the quadrupole one:
\begin{equation}
{\displaystyle \int} dx\ b_1 (x)
         =
             {\displaystyle \lim_{t\rightarrow 0}}
              -{5 \over 3} {t \over {4M^2}} F_Q (t)
           +{1 \over 9} \delta Q_{sea} 
\ \ \ ,
\label{eqn:SUM}
\end{equation}
where the sea-quark tensor polarization is defined by
$\delta Q_{sea}= {\displaystyle \int} dx  
    [8 \delta \bar u +2 \delta \bar d +\delta s +\delta \bar s ]$,
and $F_Q(t)$ is the quadrupole form factor in the unit of $e/M^2$.
Equation (\ref{eqn:SUM}) is analogous to the Gottfried integral
${\displaystyle \int} dx\ [F_2^p(x)-F_2^n(x)]
         = 1/3 + (2/3) {\displaystyle \int} dx [\bar u(x)-\bar d(x)]$.
As the Gottfried sum 1/3 is obtained by assuming the flavor
symmetry $\bar u=\bar d$, the $b_1$ sum 
\begin{equation}
{\displaystyle \int} dx\ b_1 (x)=
             {\displaystyle \lim_{t\rightarrow 0}}
              -{5 \over 3} {t \over {4M^2}} F_Q (t)=0
\ \ \ ,
\end{equation}
is obtained if there is no sea-quark tensor polarization.
Because of this assumption on the tensor polarization,
the $b_1$ sum rule is considered as 
one of the sum rules in a naive parton model.
It is similar to the Gottfried sum rule in this sense.
At this stage, all model calculations for $b_1(x)$ 
satisfy the sum rule ${\displaystyle \int} dx~ b_1(x) =0$. 
As the breaking of the Gottfried sum rule became
an interesting topic recently, it is 
worth while investigating a possible mechanism to produce
the tensor polarization $\delta Q_{sea}$, which
breaks the sum rule.

Even though the sum-rule value is expected to be 
zero for the $b_1$, it does not mean that $b_1$ itself
is zero. In fact, $b_1$ can be negative in a certain $x$ region.
For example, the deuteron $b_1$ is given by
a helicity amplitude for the nucleon convoluted
with a light-like momentum distribution of the nucleon
\cite{B1PAPER,B1D}:
\begin{eqnarray}
& & b_1^D(x)= \int_x^2 {{dy}\over y} F_1(x/y)
          \int d^3 p \left [ - {3 \over{4\pi\sqrt{2}}} 
                                \sin\alpha \cos\alpha u_s(p) u_d(p)
                          + {3 \over{16\pi}} \sin^2\alpha u_d^2(p) \right ]
     \nonumber \\
& & \ \ \ \ \ \ \ \ \ \ \ \ \ \ \ \ 
        \times \   (3\cos^2\theta-1)\left[ 1+{{p\cos\theta}\over M}
                                    +{{p^2}\over{4M^2}} \right]
           \delta\left[ y -{{p\cos\theta+E(p)}\over M} \right]
\ \ \ ,
\end{eqnarray}
where $sin \alpha$ is the D-state admixture, and $u_{s,d}$ are
S and D-wave deuteron wave functions.
The first term is due to the S-D interference
and the second to the D-state.
This is just an example of nuclear structure aspects.
Measured deuteron $b_1$ does not have to agree with the
above estimate. If a deviation from nuclear contributions
is found, $b_1$ provides important clues to physics
of non-nucleonic components in nuclei and to new
tensor structure on the parton level.

\section*{{\bf Acknowledgment}}

\vspace{-0.3cm}
This research was partly supported by the Grant-in-Aid for
Scientific Research from the Japanese Ministry of Education,
Science, and Culture under the contract number 06640406.

$~~~$

\vspace{-0.3cm}
\noindent
{* Email: kumanos@cc.saga-u.ac.jp. 
   Information on his research is available}  \\

\vspace{-0.6cm}
\noindent
{at http://www.cc.saga-u.ac.jp/saga-u/riko/physics/quantum1/structure.html.} \\

\vspace{-0.6cm}


\end{document}